\documentstyle[psfig]{mn}

\newcommand{\etal}{et al.}       
\newcommand{\HI}{H\,{\sc i}}       
\newcommand{\HII}{H\,{\sc ii}}       
\newcommand{\kms}{\,km\,s$^{-1}$}       
\newcommand{\Msun}{\,M$_{\sun}$}
\newcommand{\Lsun}{\,L$_{\sun}$}
\newcommand{\dms}[3]{{#1}\degr\,{#2}\arcmin\,{#3}\arcsec}

\newcommand{\hmfs}[4]{{#1}$^{\rm h}$\,{#2}$^{\rm m}$\,{#3}\fs{#4}}   

\title[ATCA HI Observations of the NGC~6845 Galaxy Group]
      {ATCA\thanks{The Australia Telescope Compact Array is part of the 
       Australia Telescope which is funded by the Commonwealth of Australia 
       for operation as a National Facility managed by CSIRO.}
       \HI\ Observations of the NGC~6845 Galaxy Group}
\author[Gordon \etal]
{Scott~Gordon$^1$, B\"arbel~Koribalski$^2$, Keith~Jones$^1$ \\
$^1$Department of Physics, University of Queensland, St. Lucia,
    Brisbane, QLD 4072, Australia \\
$^2$Australia Telescope National Facility, CSIRO,
    P.O. Box 76, Epping, NSW 1710, Australia }

\date{Received date; 7 March 2003}

\begin{document}
\maketitle

\begin{abstract}
We present the results of ATCA \HI\ line and 20-cm radio continuum 
observations of the galaxy Quartet NGC~6845. The \HI\ emission extends over 
all four galaxies but can only be associated clearly with the two spiral 
galaxies, NGC~6845A and B, which show signs of strong tidal interaction. 
We derive a total \HI\ mass of at least $1.8 \times 10^{10}$\Msun, most of 
which is 
associated with NGC~6845A, the largest galaxy of the group. We investigate 
the tidal interaction between NGC~6845A and B by studying the kinematics 
of distinct \HI\ components and their relation to the known \HII\ regions. 
No \HI\ emission is detected from the two lenticular galaxies, NGC~6845C and 
D. A previously uncatalogued dwarf galaxy, ATCA~J2001--4659, was detected 
4\farcm4 NE from NGC~6845B and has an \HI\ mass of $\sim 5 \times 10^{8}$\Msun.
No \HI\ bridge is visible between the group and its newly detected companion.

Extended 20-cm radio continuum emission is detected in NGC~6845A and B as
well as in the tidal bridge between the two galaxies. We derive star 
formation rates of 15 to 40\Msun\,year$^{-1}$.
\end{abstract}

\begin{keywords}
galaxies: individual: NGC~6845 --- galaxies: interactions
\end{keywords}

\section{Introduction} 
The \HI\ properties, tidal features and star formation rates in interacting 
galaxies are important to study because they reveal how various components 
of the ISM evolve under the forces of gravity. Compact groups of galaxies 
are particularly interesting because they often display a multitude of tidal 
features, which outline their interaction history, as well as high star 
formation rates. A prominent example is Stephen's Quintet (actually a galaxy 
quartet plus one foreground galaxy) which contains an \HI\ mass of 
$\sim10^{10}$\Msun\ located in several \HI\ clouds and large tidal tails 
which are clearly displaced from the disks of the four interacting galaxies 
(Williams, Yun \& Verdes-Montenegro 2002). The NGC~6845 Quartet, which is the 
focus of this paper, is similar to Stephen's Quintet in terms of distance, 
angular size, number of group members and \HI\ mass. But its \HI\ distribution 
and kinematics are very different, as we will show. Generally, a wide range 
of morphologies and kinematics is found in Hickson compact groups (Huchtmeier 
1997; Verdes-Montenegro et al. 2001), indicating the high complexity of the 
on-going interactions in these dense groups.

This paper is the third in a series of \HI\ studies of interacting galaxies,
which started with a detailed study of the spectacular interacting galaxy 
pair NGC~4038/9 ("The Antennae"; Gordon, Koribalski \& Jones 2001, see also 
Hibbard et al. 2001), followed by the peculiar galaxy system IC~2554 
(Koribalski, Gordon \& Jones 2003). Here we study the \HI\ gas distribution
and kinematics of the galaxies in the NGC~6845 group. In all three systems 
we detected additional dwarf members which is quite common in studies of 
(interacting) galaxy groups.

The NGC~6845 Quartet (Klemola~30, ESO284--IG008) is a compact group of two 
spiral and two lenticular galaxies which were first described in a proper 
motion survey by Klemola (1969) and later by Graham \& Rubin (1973). All four 
galaxies have similar systemic velocities, suggesting they are physically 
close, and are classified as peculiar. For a summary of some optical properties
see Table~1. Deep optical images (Rose 1979; Rose \& Graham 1979) show clear 
signs of galaxy interactions such as a bright tail or bridge as well as 
numerous faint wisps and filaments. The four galaxies lie roughly in a line 
(see Fig.~1) and are named B, A, C, and D from NE to SW, with NGC~6845A being 
the largest galaxy of the group. 

Optically, the NGC~6845 Quartet extends over $\sim4\arcmin \times 2\arcmin$
and has a mean optical recession velocity of $\sim$6600\kms. Assuming H$_0$
= 75\kms\,Mpc$^{-1}$ this velocity implies a Hubble distance of $D$ = 88\,Mpc
and a scale size of $1\arcmin \equiv 25.6$\,kpc. The projected size of the 
group is $\sim$100\,kpc. The two spiral galaxies, NGC~6845A and B, show the 
strongest signs of interaction, with the most prominent feature being an 
apparent bridge between the galaxies (Fig.~1; see also Rodrigues \etal\ 1999). 

NGC~6845A is a nearly edge-on, barred spiral galaxy with a bright nuclear
region as well as \HII\ regions at both ends of the bar. The latter are the 
bases of the spiral or tidal arms. Deep optical images (see e.g. Rose 1979; 
Laustsen, Madsen \& West 1987, Plate 85) clearly show the bright eastern arm 
extending towards NGC~6845B, resembling a bridge between the two galaxies. 
The much less prominent western arm shows very faint extensions toward the SW,
then appears to loop back towards NGC~6845A, passing the slightly more distant
lenticular galaxies NGC~6845C and D in projection (see also Rose \& Graham 
1979). NGC~6845B is a compact, nearly face-on spiral galaxy with a rather 
disturbed appearance. It shows several filamentary extensions which are 
likely distorted spiral arms or material pulled out of the disk.

Rodrigues \etal\ (1999) carried out extensive optical spectroscopy and 
photometry of the NGC~6845 Quartet as well as short VLA \HI\ observations. 
They found two \HI\ components associated with NGC~6845A and B, with a total
\HI\ mass of $1.5 \times 10^{10}$\Msun, as well as five very bright \HII\ 
regions in the tidal bridge. Rodrigues \etal\ also present preliminary results 
from N-body simulations of the group, which slightly favour a collision of 
NGC~6845A and B to reproduce the tidal bridge.

In the following we describe our observations and data reduction. In Section~3
we present the \HI\ properties of the NGC~6845 group as well as its radio 
continuum emission. In Section~4 we discuss the \HI\ dynamics and galaxy 
interactions. Star formation rates are analysed in Section~5 followed by the 
conclusions in Section~6. In the Appendix we show a HIPASS spectrum of 
NGC~6845 and a 20-cm radio continuum image of the radio galaxy PKS\,1955--470. 

\section{Observations \& Data Reduction} 
\HI\ observations of the NGC~6845 group were made during 1997 and 1998, using
the Australia Telescope Compact Array (ATCA). Data reduction was performed 
using the {\sc miriad} package except for the use of the {\sc AIPS} task {\sl 
momnt}. The observations were obtained in the 375, 750A, 750D, 1.5A and 6B 
arrays, for $\sim$12 hours each. Each run alternated between 30 minutes 
on-source and 3 minutes on the phase and flux calibrator PKS~1934--638. A 
16 MHz-wide bandpass was centered on 1389 MHz, returning two polarisations 
in each of the 512 channels. During calibration, highly attenuated channels 
at the bandpass edges were removed, leaving a bandwidth of $\sim$13.6 MHz or 
$\sim$2870\kms. Solar interference, which affected the shortest baselines 
during observations with the 750D and 375 arrays, was removed.

After continuum subtraction, \HI\ channel maps were made by 
Fourier-transformation of the {\em uv}-data, using `natural' weighting of
the visibilities to maximise sensitivity. The channel width used is 20\kms. 
The channels were `cleaned' giving an r.m.s. residual noise slightly higher
than the theoretical value of 0.7 mJy\,beam$^{-1}$. The resulting beam is
$43\arcsec \times 36\arcsec$. The AIPS task {\sl momnt} was applied to the 
\HI\ cube to produce integrated \HI\ intensity (0.\,moment) maps over various 
velocity ranges (see Section~3.1), an intensity-weighted mean velocity field 
(1.\,moment) and an \HI\ velocity dispersion (2.\,moment) map, using 3-pixel 
`boxcar' spatial smoothing, 3-channel Hanning smoothing and a lower flux 
density cutoff of 2.25 mJy\,beam$^{-1}$. 
Additional \HI\ moment maps from individual and combined \HI\ components 
(see Section~3.1) have been prepared by masking the channel maps.
These were smoothed by convolution with an 18\arcsec\ Gaussian, and then
clipped at the 1.5 mJy\,beam$^{-1}$ level to suppress noise. The resulting 
beam is $\sim47\arcsec \times 40\arcsec$, slightly larger than before.


The 20-cm radio continuum maps were produced using `robust' weighting
(robust = 0.5) of the line-free channels resulting in a beam size of 
$17\farcs7 \times 15\farcs8$ and an r.m.s. noise level of $\sim$0.16 
mJy\,beam$^{-1}$, higher than the theoretical value of 0.07 mJy\,beam$^{-1}$.

\section{Results} 
In Section~3.1 we describe the \HI\ components in the NGC~6845 group, 
followed by a study of the derived \HI\ moments in Section~3.2. The newly 
detected companion to the group is described in Section~3.3. Results from
the 20-cm radio continuum measurements are presented in Section~3.4. 

\subsection{The \HI\ Components} 
The \HI\ channel maps (Fig.~2) show emission in the velocity range from 
$\sim$6160 to 6740\kms, with the possibility of additional faint emission 
at higher velocities, $\sim$6840 to 6860\kms. The structure and kinematics 
of the \HI\ gas appear very complex with numerous disconnected emission 
components distributed throughout the whole velocity range. This is also well
illustrated in the \HI\ position-velocity ({\em pv}) diagram (Fig.~3).
The largest continuous \HI\ complex is clearly associated with NGC~6845A. 
The \HI\ emission coinciding with NGC~6845B is in general fainter and appears 
rather irregular. No \HI\ emission is detected from the lenticular galaxies 
NGC~6845C and D which have systemic velocities of 6816 and 7070\kms\ (see 
Table~1). 

To analyse the complex \HI\ emission we distinguish five regions which,
individually, appear continuous in position and velocity. Regions~1--4 cover 
the galaxies NGC~6845A and B, while Region~5 corresponds to the newly detected 
companion. The \HI\ properties of these regions are listed in Table~2. 
Region~1 corresponds mostly 
to NGC~6845A ($v_{\rm opt}$ = 6356\kms) and extends over by far the largest
velocity range, from 6160 to 6740\kms. It comprises most of the \HI\ gas, 
covering the disk of NGC~6845A as well as most of the arms, with a 
diameter of $\sim$3\farcm2 (83 kpc). Regions~2--4 are, in projection, close to 
NGC~6845B and parts of the tidal bridge: Region~2 (6160 to 6220\kms) has a
very narrow velocity range and covers the extended emission 
NE of NGC~6845B, whereas Region~3 (6260 to 6500\kms) appears to coincide 
mostly with the tidal bridge between NGC~6845A and B, including the tip of 
the bridge which is very close to the nucleus of NGC~6845B; Region~4 (6560 to 
6740\kms, possibly up to 6860\kms) covers numerous irregular features closely 
related to NGC~6845B ($v_{\rm opt}$ = 6753\kms). Well separated from the main 
\HI\ complex lies Region~5 (6660 to 6680\kms) which corresponds to the newly 
detected companion at $\alpha,\delta$(J2000) = \hmfs{20}{01}{15}{8}, 
\dms{--46}{59}{35}. We refer to this galaxy as ATCA~J2001--4659 after its 
observed \HI\ center position.

Fig.~3 shows the \HI\ {\em pv}-diagram along the major axes of NGC~6845A. 
Because its position angle ($PA$ = 63\degr) is close to that of the Quartet 
as a whole, all four components (Regions~1--4) are visible: the bright, 
continuous \HI\ emission corresponds to Region~1; the three fainter, but 
discontinuous complexes correspond to Regions~2--4.

\subsection{Gas distribution and kinematics} 
Fig.~4 shows the \HI\ distribution of the NGC~6845 group including the new
companion, ATCA~J2001--4659. The latter is described in Section~3.3. No \HI\ 
bridge was detected between NGC~6845 and its companion, but we notice that 
the north-eastern part of the \HI\ distribution points in its direction. 
In the following we concentrate on the \HI\ kinematics within the NGC~6845 
Quartet.

In Fig.~5 we present the moment maps of the NGC~6845 Quartet. The \HI\ 
distribution (Fig.~5a) is elongated and nearly covers all four optical 
galaxies with further extensions towards the north-east and south-west
(diameter $\sim$5\arcmin\ = 130 kpc). The \HI\ velocity field (Fig.~5b,c) 
is consistent with all the \HI\ emission coming from NGC~6845A and B, with 
NGC~6845A being the dominant, most massive galaxy. NGC~6845C and D, which 
were not detected in the \HI\ cube, have position angles nearly 90\degr\ 
different from NGC~6845A. There is no indication of any disturbance in that 
part of the velocity field. Both the \HI\ velocity field and the velocity 
dispersion (Fig.~5d) show major disturbances near NGC~6845B whereas the 
disk and south-western extension of NGC~6845A appear rather regular.

The \HI\ systemic velocity of NGC~6845A is difficult to determine because 
the overall \HI\ velocity field (as well as that of Region~1 alone) does not 
resemble that of a normal rotating spiral galaxy. Galaxy interactions affect
the gas dynamics strongly such that particularly the outer parts of the galaxy 
appear more like streaming motions than rotation. Using the major-axes 
{\em pv}-diagram of NGC~6845A (see Fig.~3) we can derive a mean systemic 
velocity of 6400--6500\kms, slightly higher than the optical velocity of 
6356\kms\ (de Vaucouleurs \etal\ 1991). Rodrigues \etal\ (1999) find an 
H$\alpha$ velocity of 6399\kms\ for the nucleus of NGC~6845A (note that their 
velocity resolution is $\sim$8\AA). The {\em pv}-diagram also highlights the 
asymmetric distribution of the \HI\ gas; the \HI\ maximum corresponds to a 
velocity of $\sim$6500\kms. For the nucleus of NGC~6845B Rodrigues \etal\ find 
$\sim$6746\kms, similar to the value quoted by de Vaucouleurs \etal\ (1991). 
But the H$\alpha$ velocities of the nearby \HII\ regions\footnote{The 
numbering and velocities of the \HII\ regions are taken from Table~6 in 
Rodrigues \etal\ (1999).} 4 and 5 (6314 and 6201\kms, respectively) which 
they associate with the tip of the tidal bridge, are very different. Using 
position and velocity information we can associate their \HII\ region~5 with 
our \HI\ Region~2, \HII\ regions~1--4 with \HI\ Region~3, and the nucleus 
of NGC~6845B with \HI\ Region~4. Because of its irregular emission, we cannot 
derive a meaningful \HI\ systemic velocity for NGC~6845B.

Fig.~6 shows \HI\ spectra for the NGC~6845A/B system and the newly detected
companion, respectively. The multi-component spectrum of NGC~6845A/B, which 
covers a velocity range of 580\kms, is typical for a disturbed system. 

\subsection{The companion ATCA~J2001--4659} 
The galaxy ATCA~J2001--4659 has an \HI\ systemic velocity of $6667 \pm 
6$\kms, lies 4\farcm4 north of the NGC~6845B, and is a new member of 
the NGC~6845 galaxy group (see Fig.~4). It is slightly resolved in \HI\ 
with a diameter of $\sim55\arcsec \times 40\arcsec$ and a position angle
of $PA$ = 40\degr. Its optical counterpart (see Fig.~7) has a diameter of 
$\sim$20\arcsec, less than half the \HI\ diameter. The narrow velocity 
width of ATCA~J2001--4659 ($w_{50} = 40 \pm 12$\kms; see Fig.~6b) together 
with its optical appearance suggest that it is a previously uncatalogued 
dwarf galaxy. We measure an integrated \HI\ flux density of $0.3 \pm 0.1$ 
Jy\kms\ and derive an \HI\ mass of 5 ($\pm$ 2) $\times 10^8$\Msun.
ATCA~J2001--4659 appears undisturbed both in \HI\ and optical images.

No \HI\ bridge was detected between ATCA~J2001--4659 and the NGC~6845 Quartet. 
We note that the new companion lies roughly on the same NE-SW line as the 
galaxies NGC~6845A--D. NGC~6845B, which is closest in velocity and position 
to ATCA~J2001--4659, shows rather irregular behaviour including a faint 
optical and \HI\ extension towards it.

\subsection{Radio Continuum Emission} 
We detected extended 20-cm radio continuum emission in the two interacting 
spirals NGC~6845A and B, as well as the tidal bridge, with a flux density 
of $37 \pm 2$ mJy (Fig.~10). No radio continuum emission was detected in the 
lenticular galaxies NGC~6845C and D. 

The emission in NGC~6845A ($\sim$26 mJy) is concentrated along the bar 
with a bright source at its north-eastern end. The 20-cm peak flux of 
6.6 mJy\,beam$^{-1}$ occurs at $\alpha,\delta$(J2000) = \hmfs{20}{00}{59}{4},
\dms{--47}{04}{09}, NE of the nucleus, roughly coincident with the bright 
\HII\ region~7 (L$_{H\alpha} = 2.2 \times 10^{40}$\,ergs\,s$^{-1}$, Rodrigues 
\etal\ 1999). Extended, but weaker radio continuum emission covers the 
remaining part of the bar, including the nucleus and \HII\ region~8. Both
are a factor 2--3 fainter than \HII\ region~7. The total H$\alpha$ luminosity
in NGC~6845A is at least $3.8 \times 10^{40}$\,ergs\,s$^{-1}$.

The emission close to NGC~6845B ($\sim$11 mJy) is extended and covers the
nucleus of NGC~6845B as well as the \HII\ regions~4 and 5, which are 
located in the tip of the tidal bridge. Their H$\alpha$ luminosities are 
11, 9.1, and $6.5 \times 10^{39}$\,ergs\,s$^{-1}$, respectively (Rodrigues
\etal\ 1999).
We find a 20-cm peak flux of 2.8 mJy\,beam$^{-1}$ at $\alpha,\delta$(J2000) 
= \hmfs{20}{01}{05}{6}, \dms{--47}{03}{36}), coincident with the nucleus of
NGC~6845B (see Table~1). 

The 20-cm radio continuum flux densities quoted for NGC~6845A and B both 
contain a small amount of radio continuum emission from the tidal bridge. 
Although faint, the radio continuum bridge appears slightly offset to the 
North from the optical bridge, and barely covers the \HII\ regions 1--3 
(Rodrigues \etal\ 1999). The latter are rather faint and have a combined 
H$\alpha$ luminosity of $7 \times 10^{39}$\,ergs\,s$^{-1}$. In general, we 
find the strength of the radio continuum emission correlates well with the 
H$\alpha$ luminosity. 

\section{Discussion} 
The integrated \HI\ flux density of NGC~6845A and B together is at least
$10 \pm 1$ Jy\kms\ (see also Appendix A) resulting in an \HI\ mass of 
$1.8 \times 10^{10}$\Msun, about 20\% higher than the value derived by 
Rodrigues et al. (1999) who appear to miss \HI\ flux at velocities below 
6350\kms. We use $M_{\rm HI}$ = $2.36~10^5 D^2 F_{\rm HI}$, where $D$ = 
88\,Mpc is the adopted distance to NGC~6845 and $F_{\rm HI}$ is the 
integrated \HI\ flux density in Jy\kms. The \HI\ mass of NGC~6845A/B 
comprises about 6\% of its total Keplerian mass of 
$\sim3 \times 10^{11}$\Msun\ (Rodrigues \etal\ 1999). 

There are many indications both in optical images and our \HI\ data that 
NGC~6845A and B are strongly interacting, whereby the smaller galaxy, 
NGC~6845B, has been severely disrupted. The two arms of NGC~6845A could 
either be distorted spiral arms, like in NGC~6872 (Horellou \& Koribalski
2002) or tidal tails, like in NGC~4038/9 (Gordon et al. 2001). 

In the following we discuss possible scenarios to explain the four \HI\
components in the NGC~6845 Quartet. We have shown that Region~1 is clearly 
associated with NGC~6845A and Region~4 with NGC~6845B. The \HI\ Regions~2 
\& 3 lie towards the northeast of the NGC~6845A disk and are likely tidal 
material. We investigate if Regions~2 \& 3 are mainly associated with 
NGC~6845A or with NGC~6845B.

Fig.~8 shows separate \HI\ moment maps of Region~1, Regions~2+3, and Region~4. 
By analysing the individual components we gain some understanding of the 
abrupt velocity changes and high velocity dispersion in the north-eastern part 
of the NGC~6845 Quartet (see Fig.~4). Region~1 has by far the highest \HI\ 
mass, followed by Regions~2+3 and Region~4 with $M_{\rm HI}$ = 13.6, 3.9, and 
$1.8 \times 10^{9}$\Msun, respectively, (see Table~2). The velocity field of 
Region~4 (NGC~6845B) is rather irregular in contrast to the other components.

First we suggest that Regions~2+3 are mainly associated with 
NGC~6845A, as also discussed by Rodrigues \etal\ (1999). This would imply 
that NGC~6845A is a rather symmetric galaxy, where Region~1 includes the 
central disk and the western spiral arm, and Regions~2+3 make up the eastern 
spiral arm and the extended tidal bridge. Fig.~9 shows the \HI\ moment maps 
of the combined Regions~1--3, which have a total \HI\ mass of $1.7 \times 
10^{10}$\Msun. The western spiral arm is optically very faint, whereas the 
\HI\ is bright and extends just beyond NGC~6845D. The eastern spiral arm 
is elongated and appears as an optically very bright tidal bridge which, in 
this scenario, continues past NGC~6845B towards the north-east. This is 
strongly supported by the velocity gradient both in \HI\ (see Fig.~9b) and 
H$\alpha$ (Rodrigues \etal\ 1999). The \HI\ gas extends slightly further 
than the north-eastern most tip of the optical tidal bridge. The velocity 
field of Regions~1--3 shows a rather smooth velocity gradient across the
\HI\ distribution with only a small discontinuity around 6350\kms. The 
velocity range from 6250 to 6450\kms\ is characterised by the apparent 
merging of Regions~1 and 3 (see Fig.~2). It is also the location of \HII\ 
regions 1--3 in Rodrigues \etal\ (1999) and thus recent star formation.

Since Regions~2--4 are located close to NGC~6845B, there is a possibility
that Regions~2+3 (see Fig.~8, middle panel) correspond to tidal tails 
emanating from NGC~6845B, ending $\ge$1\farcm6 or 42\,kpc north-east of 
the center at low velocities near 6170\kms. In that case, the \HI\ gas 
associated with NGC~6845B covers a velocity range of $\sim$560\kms,
comparable to that of NGC~6845A, and has an \HI\ mass of about $6 \times 
10^9$\Msun. 

\section{Star Formation Rates} 
To estimate the star formation rate (SFR) of NGC~6845 from its 20-cm radio
continuum flux density we use 
SFR (M$_{\sun}$\,year$^{-1}$) = 0.14 $D^2$ F$_{\rm 20cm}$
derived from Condon (1992) and Haarsma \etal\ (2000), where $D$ is the 
distance in Mpc and $F_{\rm 20cm}$ the 20-cm radio continuum flux density
in Jy. For NGC~6845A ($D$ = 88\,Mpc) we measure $F_{\rm 20cm}$ = 26 mJy
and hence SFR = 28\Msun\,year$^{-1}$. For NGC~6845B ($F_{\rm 20cm}$
= 11 mJy) we calculate SFR = 12\Msun\,year$^{-1}$.

The star formation rate of a galaxy can also be estimated from its 
far-infrared luminosity, $L_{\rm FIR}$, using SFR (M$_{\sun}$\,year$^{-1}$) 
= 0.17 $L_{\rm FIR}$ (Kennicutt 1998), with $L_{\rm FIR}$ in units of 
10$^9$\Lsun. Using a FIR luminosity of $8.5 \times 10^{10}$\Lsun\
for the NGC~6845 group (Rodrigues \etal\ 1999) we calculate a SFR of
15\Msun\,year$^{-1}$. Solomon \& Sage (1988) report average FIR 
luminosities for a range of galaxy types: for non-interacting galaxies 
they find $L_{\rm FIR}$ = $1.1 - 3.0 \times 10^{10}$\Lsun, for galaxies 
with tidal tails and bridges, like NGC~6845, they find much higher values, 
but also larger variations: 
$L_{\rm FIR} = 16^{+20}_{-9} \times 10^{10}$\Lsun\ 
or SFRs of 12 -- 61\Msun\,year$^{-1}$. 

Unlike the far-infrared, \HII\ and UV, the radio flux density traces both 
obscured and unobscured star formation. Following Helou, Soifer \&
Rowan-Robinson (1985) we find $q \approx 2.4$ for the ratio of the 
infrared to radio emission in NGC~6845. This is close to the mean value 
found by Condon, Anderson \& Helou (1991) implying that both the radio 
and FIR luminosities originate in the same population.

\section{Conclusions} 
The two interacting spiral galaxies, NGC~6845A and B, as well as the tidal 
bridge were detected in both \HI\ and the 20-cm radio continuum emission.
NGC~6845A is \HI-rich (M$_{\HI} \ga\ 10^{10}$\Msun) and covers a wide 
velocity range of 580\kms. NGC~6845B is smaller than NGC~6845A and has
an \HI\ mass of at least $\sim2 \times 10^{9}$\Msun. Neither \HI\ nor radio 
continuum emission was detected in the two lenticular galaxies, NGC~6845C 
and D. Both optical images and our \HI\ data confirm that NGC~6845A and B 
are strongly interacting. The star formation rates in both galaxies are 
enhanced and similar to those in other interacting galaxies. NGC~6845B, the 
smaller and less-massive galaxy, appears to have been severely disrupted.

We discovered a dwarf companion, ATCA~J2001--4659 ($v_{\rm HI} = 6667 
\pm 6$\kms), to the NGC~6845 group, located about 4\farcm4 NE away from 
NGC~6845B. It has an \HI\ mass of 5 ($\pm$ 2) $\times 10^{8}$\Msun\ and 
is slightly resolved. 

There is no evidence that NGC~6845C and D have interacted with each other 
or with the other group members. Their optical images show no obvious 
distortions apart from a possible warping of the northern tip of NGC~6845C. 
No irregularities in the \HI\ velocity field are seen in their vicinity.

Extended radio continuum emission was detected in NGC~6845A and B as well as
the tidal bridge with a total 20-cm flux density of $37 \pm 2$ mJy. NGC~6845A 
shows extended emission roughly along the bar with the brightest knot towards 
the NE, coinciding with the \HII\ region~7 (see Rodrigues \etal\ 1999). 
No continuum emission was detected in NGC~6845C and D. 

As noted in the introduction, the NGC~6845 Quartet has some similarity to 
Stephen's Quintet. But the \HI\ distribution in both compact galaxy groups 
is very different: whereas there is no \HI\ gas detected within the disks of 
the four interacting galaxies in Stephen's Quintet (Williams et al. 2002), 
the \HI\ gas in the NGC~6845 Quartet is concentrated in one extended cloud 
enveloping all four members. Projection effects may play a role as the
four interacting galaxies in Stephen's Quintet are close to face-on whereas
the galaxies in the NGC~6845 Quartet are nearly edge-on. The shape of the \HI\
tidal tails in both groups suggests an orientation similar to that of the
respective galaxies which would mean that the \HI\ is largely concentrated 
in a plane. Another example of a compact group with predominantly edge-on
galaxies is the Hickson Compact Group (HCG) 26; the \HI\ distribution and 
velocity field in HCG~26 (Williams \& van Gorkom 1995) is remarkably similar 
to that of the NGC~6845 Quartet. In these edge-on systems it is more 
difficult to analyse the 
interactions between galaxies as the curvature of the tidal tails, which are
often projected onto the galaxies themselves, can only be inferred from
the velocity field. Both in the NGC~6845 Quartet and HCG~26, higher resolution
\HI\ observations are needed to study the galaxy interactions in detail.

Since NGC6845C and D appear to play a marginal role in the quartet, we may
compare NGC~6845A/B with other interacting galaxy pairs. E.g., we note some 
similarity between the galaxy NGC~6845A and the giant barred spiral galaxy 
NGC~6872. Both galaxies have at least one smaller companion, are very extended 
($\sim$100 kpc) and cover a large velocity range. Their extended spiral arms 
appear to be tidally stretched and both are gas-rich galaxies with \HI\ masses 
above 10$^{10}$\Msun. Their gas dynamics differ dramatically indicating 
different evolutionary stages. Whereas the \HI\ gas is centrally concentrated 
in NGC~6845, it is predominantly located in the spiral arms of NGC~6872 
(Horellou \& Koribalski 2002).

\section*{Acknowledgements}
\begin{itemize}
\item This research has made use of the NASA/IPAC Extragalactic Database 
  (NED), which is operated by the Jet Propulsion Laboratory, California 
  Institute of Technology, under contract with NASA.
\item We acknowledge the use of images from the Digitised Sky Survey (DSS) 
  based on photographic data obtained using the UK Schmidt Telescope.
\end{itemize}

\section*{Appendix A: HIPASS J2000--47}
The integrated \HI\ flux density of NGC~6845 (HIPASS J2000--47) as obtained
from the \HI\ Parkes All-Sky Survey (HIPASS) is $\sim$16$\pm$3 Jy\kms. Fig.~11
shows the Hanning smoothed HIPASS spectrum (r.m.s. noise level = 8.4 mJy), 
which has a much lower velocity resolution and sensitivity than the ATCA 
\HI\ spectrum (Fig.~6a), but shows a very similar shape (as expected). 
The horizontal line marks the fitted first order baseline, the vertical 
lines mark the velocity range of the detected \HI\ emission, and the dots 
show the locations of the peak flux density (50 mJy), 50\% velocity width 
(376\kms) and 20\% velocity width (630\kms). For details regarding HIPASS 
see Staveley-Smith et al. (1996) and Barnes et al. (2001).

\section*{Appendix B: PKS\,1955--470}
PKS\,1955--470 (PMN\,J1958--4657) is a double-lobe radio galaxy (see also
Jones \& McAdam 1992). No optical counterpart has so far been identified.
The radio continuum flux densities measured at 0.408, 0.843, 2.7 and 5.0 GHz 
are 2.0, 1.25, 0.340 and 0.25 Jy, respectively; resulting in a spectral index
of $\alpha\ = -0.5$. After primary beam correction we measure a 1.4 GHz 
integrated flux density of 0.77 Jy.

\begin{table*} 
\caption{Optical properties for the galaxies in the NGC~6845 Quartet
         (de Vaucouleurs \etal\ 1991).}
\begin{tabular}{lccccc}
\hline
Galaxy Name   & Center position & $v_{\rm opt}$ & $B_{\rm T}$ & Type & Size\\
              & RA (J2000) DEC  &  [\kms]  & [mag] & &  \\ 
\hline
NGC~6845A     & \hmfs{20}{00}{58}{1}~\dms{--47}{04}{12}
              & $6356\pm18$~~   & 13.65
	      & SBS3*P   & $3\farcm7 \times 1\farcm7$ \\
NGC~6845B     & \hmfs{20}{01}{05}{9}~\dms{--47}{03}{35}
              & $6753\pm18$~~   & 14.86
	      & SB.3?P   & $1\farcm1 \times 0\farcm6$ \\
NGC~6845C     & \hmfs{20}{00}{56}{6}~\dms{--47}{05}{02}
              & $7070\pm54^{*}$ & 16.3~~
	      & L.+*P/   & $0\farcm6 \times 0\farcm2$ \\
NGC~6845D     & \hmfs{20}{00}{53}{4}~\dms{--47}{05}{38}
              & $6816\pm44^{*}$ & 15.5~~
	      & L..*P/   & $0\farcm8 \times 0\farcm5$ \\
\hline
\end{tabular}
\flushleft
$^{*}$ The velocities of NGC~6845C and D are reversed in Rose \& Graham (1979).
\end{table*}

\begin{table*} 
\caption{\HI\ properties of the five components in the NGC~6845 group.
  Note that Regions~1--4 correspond to different parts of NGC~6845A and B; 
  Region~5 is the newly detected companion, ATCA~J2001--4659.}
\begin{tabular}{lccccccc}
\hline
~Region & $F_{\rm HI}$ & $M_{\rm HI}$ & \HI\ peak flux  & $N_{\rm HI}$
	& \multicolumn{2}{c}{peak position}      
	& velocity range \\
numbers & [Jy\kms] & [$10^9$\Msun] & [Jy\,beam$^{-1}$\kms]
        & [$10^{20}$\,cm$^{-2}$] & \multicolumn{2}{c}{RA (J2000) DEC} 
	& [\kms] \\
\hline
1   & 7.5 & 13.6 & 3.1&18.3&\hmfs{20}{00}{57}{9}&\dms{--47}{04}{12}& 6160 -- 6740 \\
2   & 0.6 & ~1.0 & 0.3&~1.6&\hmfs{20}{01}{09}{1}&\dms{--47}{03}{18}& 6160 -- 6220 \\
3   & 1.5 & ~2.8 & 0.9&~5.3&\hmfs{20}{01}{05}{0}&\dms{--47}{03}{48}& 6260 -- 6500 \\
4   & 1.0 & ~1.8 & 0.6&~3.4&\hmfs{20}{01}{06}{2}&\dms{--47}{03}{30}& 6560 -- 6740 \\
5   & 0.3 & ~0.5 & 0.2&~1.1&\hmfs{20}{01}{15}{6}&\dms{--46}{59}{30}& 6660 -- 6680 \\
\hline
2+3  &  2.1 & ~3.9 & 1.0 &~6.1 &\hmfs{20}{01}{06}{2}&\dms{--47}{03}{36}& 6160 -- 6500 \\
2--4 &  3.2 & ~5.7 & 1.6 &~9.4 &\hmfs{20}{01}{06}{2}&\dms{--47}{03}{36}& 6160 -- 6740 \\
1--4 & 10   & 18.0 & 3.1 &18.3 &\hmfs{20}{00}{57}{9}&\dms{--47}{04}{12}& 6160 -- 6740 \\
\hline
\end{tabular}
\end{table*}

\begin{figure*} 
\begin{tabular}{c}
\end{tabular}
\caption{Optical image of the NGC~6845 Quartet obtained from the Second 
 Generation Digitised Sky Survey (DSS, red-filter).}
\end{figure*}

\begin{figure*} 
\caption{\HI\ channel maps (contours) of the NGC~6845 Quartet and its 
 newly detected companion, ATCA~J2001--4659, overlaid onto an optical image
 from the DSS. Heliocentric velocities (in \kms) are given at the top left 
 corner of each plane. The beam is shown at the bottom left. The contour 
 levels are: $\pm$2.2, 3.1, 4.3, 6.1, 8.6, and 12 mJy\,beam$^{-1}$. 
 The regions 1--5 which were introduced in Section~3.1 are also labelled.}
\end{figure*}
\setcounter{figure}{1}
\begin{figure*}
\caption{ continued.} 
\end{figure*}
\setcounter{figure}{1}
\begin{figure*}
\caption{ continued.} 
\end{figure*}
\setcounter{figure}{1}
\begin{figure*}
\caption{ continued.} 
\end{figure*}

\begin{figure*} 
\begin{tabular}{ccc}
\end{tabular}
\caption{\HI\ position-velocity diagram along the major axes of NGC~6845A
  ($PA$ = 63\degr). The contour levels are: $\pm$2.2, 3.1, 4.3, 6.1, and 
  8.6 mJy\,beam$^{-1}$.}
\end{figure*}

\begin{figure*} 
\begin{tabular}{c}
\end{tabular}
\caption{Integrated \HI\ intensity map (contours) of the NGC~6845 group
 overlaid onto an optical image from the DSS. The contour levels are: 0.040, 
 0.0566, 0.080, 0.113, 0.160, 0.226, 0.320, 0.453, 0.640, 0.905, 1.280, 1.910, 
 and 2.560\,Jy\,beam$^{-1}$\kms. The beam ($43\arcsec \times 36\arcsec$) is
 shown at the bottom left.}
\end{figure*}

\begin{figure*} 
\begin{tabular}{cc}
\end{tabular}
\caption{\HI\ moment maps of the NGC~6845 Quartet.
{\bf(---a)} Integrated \HI\ intensity (contours) overlaid onto an optical 
  image from the DSS. The contour levels are 0.040, 0.0566, 0.080, 0.113, 
  0.160, 0.226, 0.320, 0.453, 0.640, 0.905, 1.280, 1.910, and 
  2.560\,Jy\,beam$^{-1}$\kms.
{\bf(---b)} The mean \HI\ velocity field. The contour levels range from 6160
  to 6680\kms\ in steps of 40\kms\ (white contours start at 6440\kms). 
  The greyscale ranges from 6100\kms\ (white) to 6800\kms\ (black). 
{\bf(---c)} The \HI\ velocity field (contours) overlaid onto an optical 
  image from the DSS. Contours as in (b).
{\bf(---d)} The mean \HI\ velocity dispersion. The contour levels are 5, 
  10, 20, 30, 60, 70, 90, 120, 150, and 180\kms. The greyscale ranges 
  from 5\kms\ (white) to 300\kms\ (black).
  The beam is shown at the bottom left of each panel.}
\end{figure*}

\begin{figure*} 
\begin{tabular}{cc}
\end{tabular}
\caption{ATCA \HI\ spectra of NGC~6845A/B (left) and the newly detected 
         companion ATCA~J2001--4659 (right).}
\end{figure*}

\begin{figure*} 
\caption{Integrated \HI\ intensity map (contours) of the galaxy 
  ATCA~J2001--4659, a newly detected companion to the NGC~6845 Quartet, 
  overlaid onto a red-filter optical image from the second-generation 
  DSS. The contour levels are: 0.040, 0.056, 0.080, 0.113, and 
  0.160\,Jy\,beam$^{-1}$\kms. The beam is shown at the bottom left.}
\end{figure*}

\begin{figure*} 
\begin{tabular}{c}
\end{tabular}
\caption{Integrated \HI\ intensity maps (left) and velocity fields (right) 
   of Region~1 (top), Regions~2+3 (middle) and Region~4 (bottom). The \HI\
   contour levels are: 0.113, 0.160, 0.226, 0.320, 0.453, 0.640, 0.905, 
   1.280, 1.910 and 2.560 Jy\,beam$^{-1}$\kms. The velocity contours range 
   from are 6160 to 6800\kms\ in steps of 40\kms\ (white contours start at 
   6440\kms). The greyscale ranges from 6100\kms\ (white) to 6800\kms\ 
   (black). The beam is shown at the bottom left of each panel 
   (here: $46\farcs6 \times 40\farcs2$).}
\end{figure*}

\begin{figure*} 
\begin{tabular}{c}
\end{tabular}
\caption{Integrated \HI\ intensity (left) and velocity field (right) of the
   combined Regions~1--3. The contour levels and greyscales are as in Fig.~8.}
\end{figure*}

\begin{figure*} 
\caption{20-cm radio continuum emission (contours) of the NGC~6845 Quartet 
   overlaid onto an optical image from the DSS. The beam of $17\farcs7 \times 
   15\farcs8$ is shown at the bottom left. The contour levels are $\pm$0.54, 
   0.76, 1.08, 1.53, 2.16, 3.05, 4.32 and 6.11 mJy\,beam$^{-1}$.} 
\end{figure*}

\begin{figure*} 
\caption{HIPASS spectrum of NGC~6845 (see Appendix A).}
\end{figure*}

\begin{figure*} 
\caption{ATCA 20-cm radio continuum image (contours) of the double-lobe radio 
   galaxy PKS\,1955--470 overlaid onto an optical image from the DSS (see 
   Appendix B).}
\end{figure*}

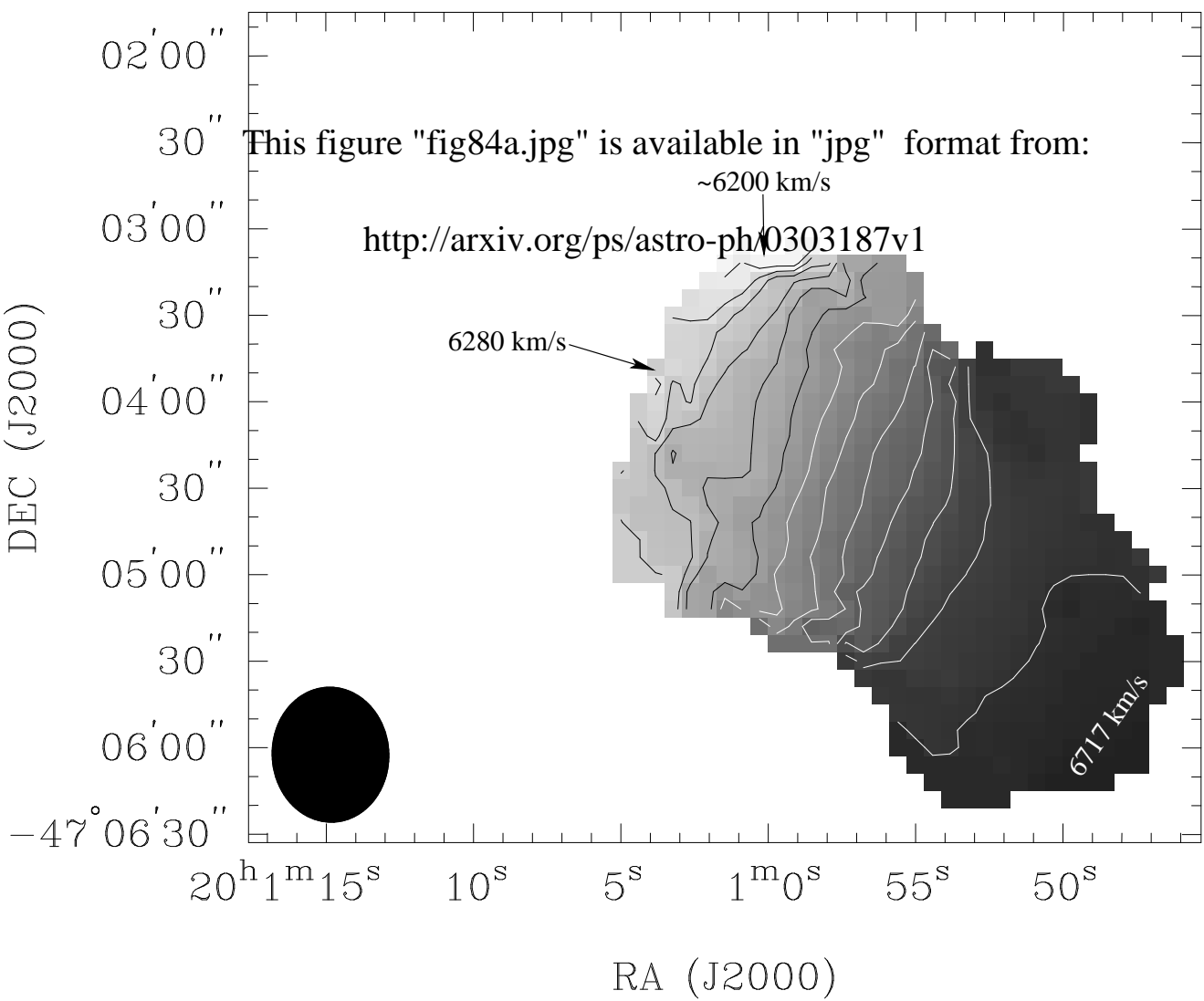
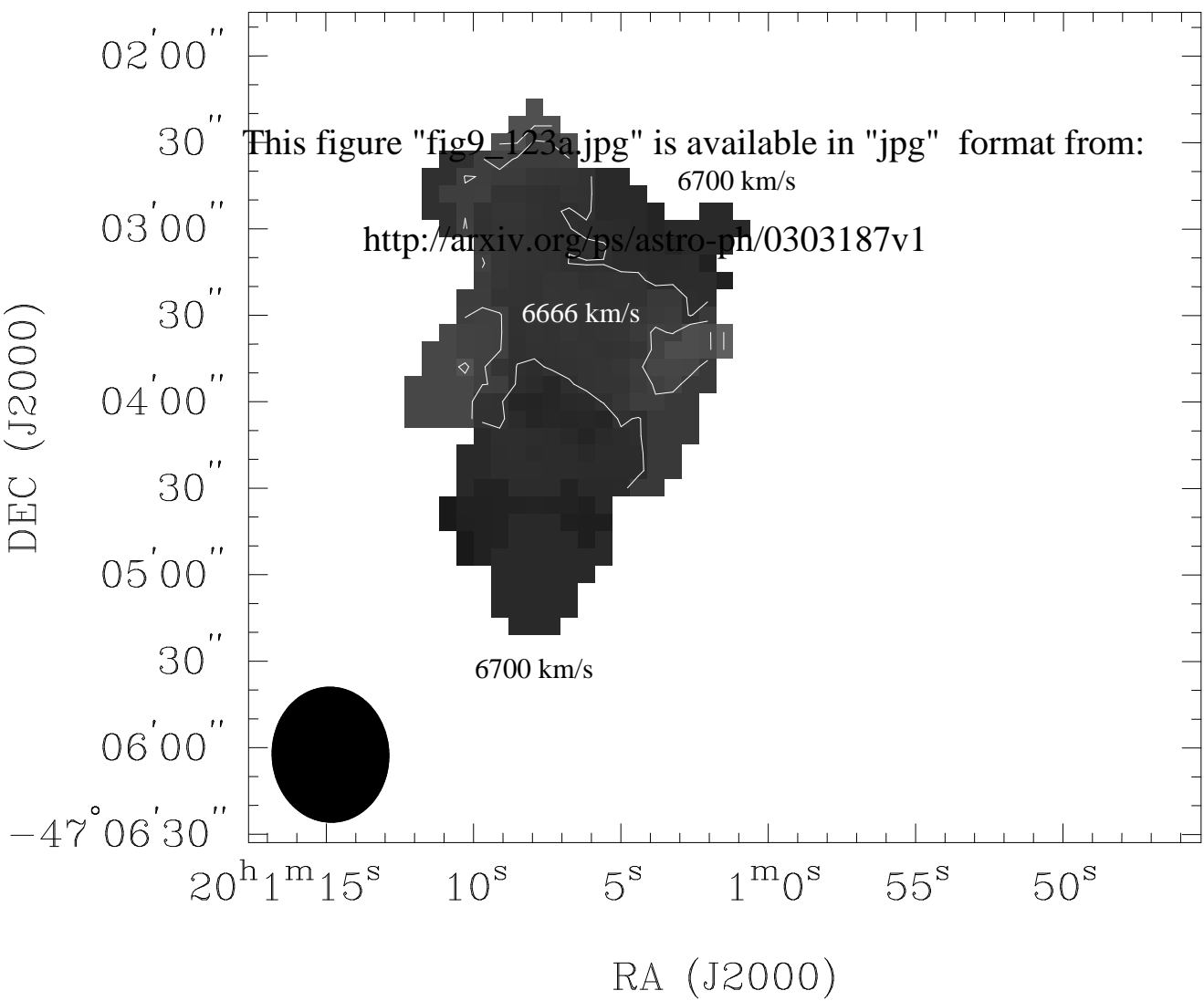
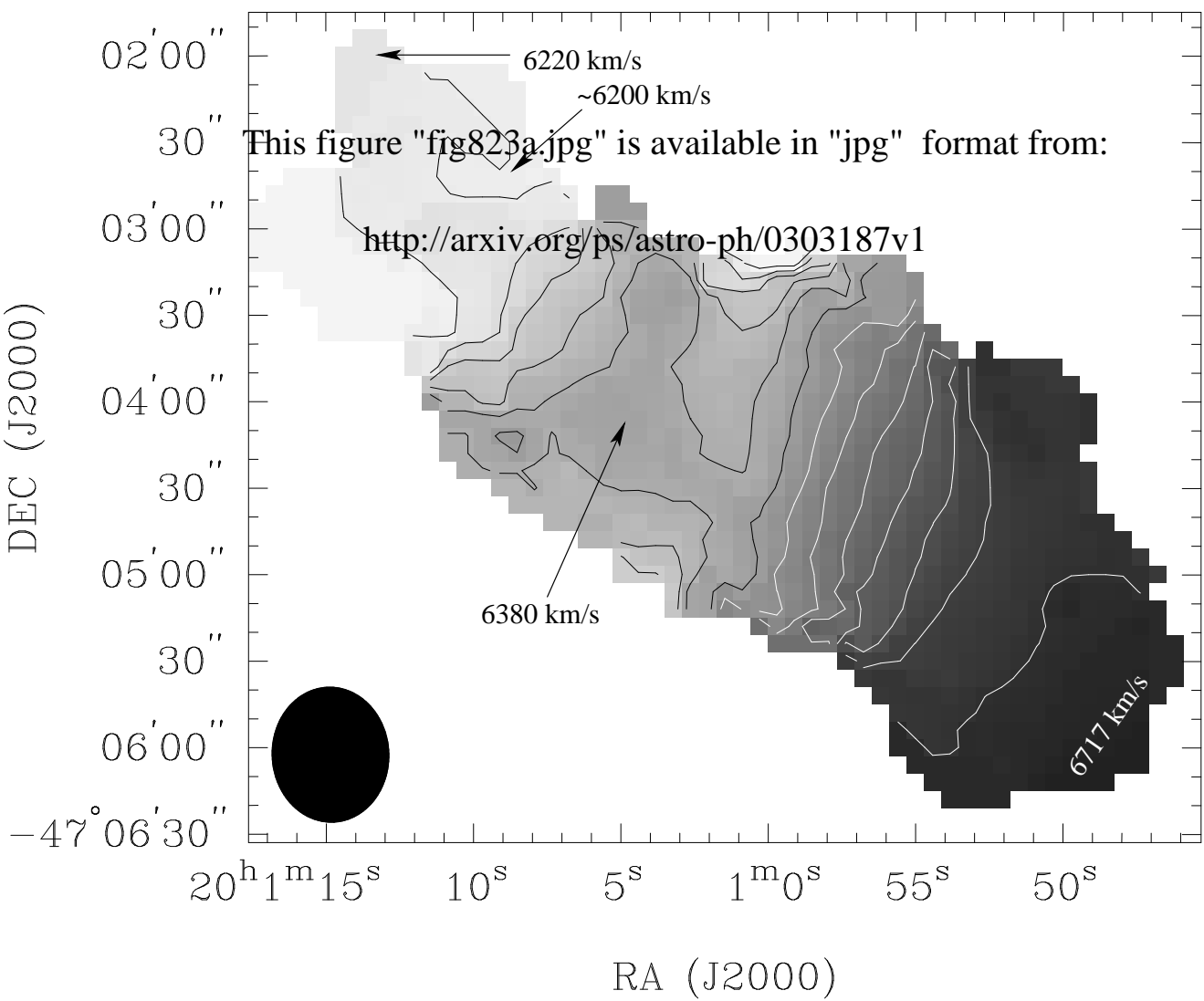
\end{document}